\documentclass{article}
\usepackage[utf8]{inputenc}
\usepackage[left=3cm,right=3cm,top=3cm,bottom=3cm]{geometry}
\usepackage{pdfpages}
\usepackage{hyperref}
\usepackage{listings}
\usepackage{xcolor}
\usepackage{setspace} 
\usepackage{tocloft}
\usepackage{amsmath}
\usepackage{chngcntr }
\usepackage[toc,page]{appendix}
\usepackage[T1]{fontenc}
\usepackage[nottoc]{tocbibind}
\usepackage[compact]{titlesec}
\usepackage[utf8]{inputenc}
\usepackage[sectionbib]{natbib}
\usepackage{chapterbib}    
\setcitestyle{aysep={,}}

\usepackage{multirow}
\usepackage{setspace}
\usepackage{array}

\newcolumntype{L}[1]{>{\raggedright\let\newline\\\arraybackslash\hspace{0pt}}m{#1}}
\newcolumntype{C}[1]{>{\centering\let\newline\\\arraybackslash\hspace{0pt}}m{#1}}
\newcolumntype{R}[1]{>{\raggedleft\let\newline\\\arraybackslash\hspace{0pt}}m{#1}}

\usepackage{subfig}
\usepackage{bm}
\usepackage{multirow}
\usepackage{tikz}
\usetikzlibrary{shapes,arrows}
\usepackage{blkarray}
\usepackage[affil-it]{authblk}

\usepackage{siunitx}

\begin{document}

\title{The impact of modelling choices on modelling outcomes: a spatio-temporal study of the association between COVID-19 spread and environmental conditions in Catalonia (Spain)}

\author[1]{Álvaro Briz-Redón}
\affil[1]{Statistics Office, City Council of València, Spain}
\affil[ ]{\normalfont alvaro.briz@uv.es}

\maketitle

\begin{abstract}

The choices that researchers make while conducting a statistical analysis usually have a notable impact on the results. This fact has become evident in the ongoing research of the association between the environment and the evolution of the COVID-19 pandemic, in light of the hundreds of contradictory studies that have already been published on this issue in just a few months. In this paper, a COVID-19 dataset containing the number of daily cases registered in the regions of Catalonia (Spain) since the start of the pandemic is analysed. Specifically, the possible effect of several environmental variables (solar exposure, mean temperature, and wind speed) on the number of cases is assessed. Thus, the first objective of the paper is to show how the choice of a certain type of statistical model to conduct the analysis can have a severe impact on the associations that are inferred between the covariates and the response variable. Secondly, it is shown how the use of spatio-temporal models accounting for the nature of the data allows understanding the evolution of the pandemic in space and time.

\end{abstract}

%\keywords{COVID-19, spatio-temporal models, environmental covariates, INLA, space-time interaction, relative risk}

\section{Introduction}

The COVID-19 pandemic has led in a few months to a very large number of related scientific outcomes. Many of the studies on the COVID-19 focus on the evolution of viral transmission, or the clinical factors that increase the risk of contagion, among other relevant topics. In particular, one of the most consolidated lines of research is dedicated to clarifying how certain environmental or meteorological factors have had an impact (or may have in the future) on the evolution of COVID-19 at a local, national, or global level.

At the time of writing (September 2020), hundreds of statistical analyses about the effect of the environment on the evolution of COVID-19 have already been published. Specifically, the influence of temperature, humidity, or solar radiation (among other variables) on the transmission of the virus has been massively investigated at a macroscopic level, considering municipalities, regions, countries, etc., as the spatial units of analysis. Surprisingly (to some extent), the results provided by these studies are sometimes very different, or even opposite, as shown by the several reviews that have been published on this topic \citep{briz2020effect,shakil2020covid,yuan2020humidity}. Some of the discrepancies found between studies could be due to the different ranges of values that the main environmental variables present depending on the area of the world being analysed, but it is also very reasonable to think that certain methodological choices such as the type of statistical model, the geographical unit of analysis, or the set of covariates also have a notable impact on the results. In relation to this fact, it is worth noting that among the studies already published on this topic, very different types of statistical and modelling techniques have been employed \citep{briz2020effect}, including correlation analyses \citep[e.g.,][]{tosepu2020correlation}, generalised additive models \citep[e.g.,][]{xie2020association}, panel data models \citep[e.g.,][]{sobral2020association}, spatio-temporal models \citep[e.g.,][]{briz2020spatio}, or epidemiological models such as the susceptible-infected-recovered-susceptible (SIRS) model \citep[e.g.,][]{baker2020susceptible}. Thus, the spatio-temporal nature of the data under analysis has been taken into account in only a relatively small percentage of studies, despite the importance of accounting for spatial and temporal patterns to explain and model the evolution of the pandemic with greater accuracy, as shown in several recent studies. For instance, \cite{guliyev2020determining} compared different panel data models and concluded that the spatially lagged X (SLX) model showed the greatest performance in modelling COVID-19 confirmed, death, and recovered rates. Moreover, \cite{mollalo2020gis} verified that geographically weighted regression models accounting for spatial heterogeneity and scale outperformed non-spatial models in modelling COVID-19 spread. Finally, in several studies developed at different levels of spatio-temporal aggregation, it has been identified that COVID-19 cases tend to be highly concentrated in space and time \citep{arauzo2020first,cordes2020spatial,desjardins2020rapid,hohl2020daily}. 

The purpose of this paper is twofold. The first objective is to highlight how certain modelling choices may affect the analysis of a spatio-temporal dataset, particularly in studying the impact of the environment on the development of the COVID-19 pandemic. To this end, the comparison starts with a rather general model without neither spatial nor temporal effects (basic generalised linear models), to which different spatial, temporal, and spatio-temporal terms are then added to properly account for the nature of the data. The second objective consists in exploring how the inclusion of spatio-temporal effects in the model can be helpful to understand the dynamics of the COVID-19 pandemic.

The paper is structured as follows. Section 2 includes a brief description of the data used for the analysis. In Section 3, the different statistical models considered for the analysis are presented. The results provided by each of the models are displayed and compared in Section 4. Finally, some concluding remarks are provided in Section 5.

\section{Data}

\subsection{Study area}

\begin{figure}[ht] 
  \centering
  \subfloat[]{\includegraphics[width=5cm,angle=-90]{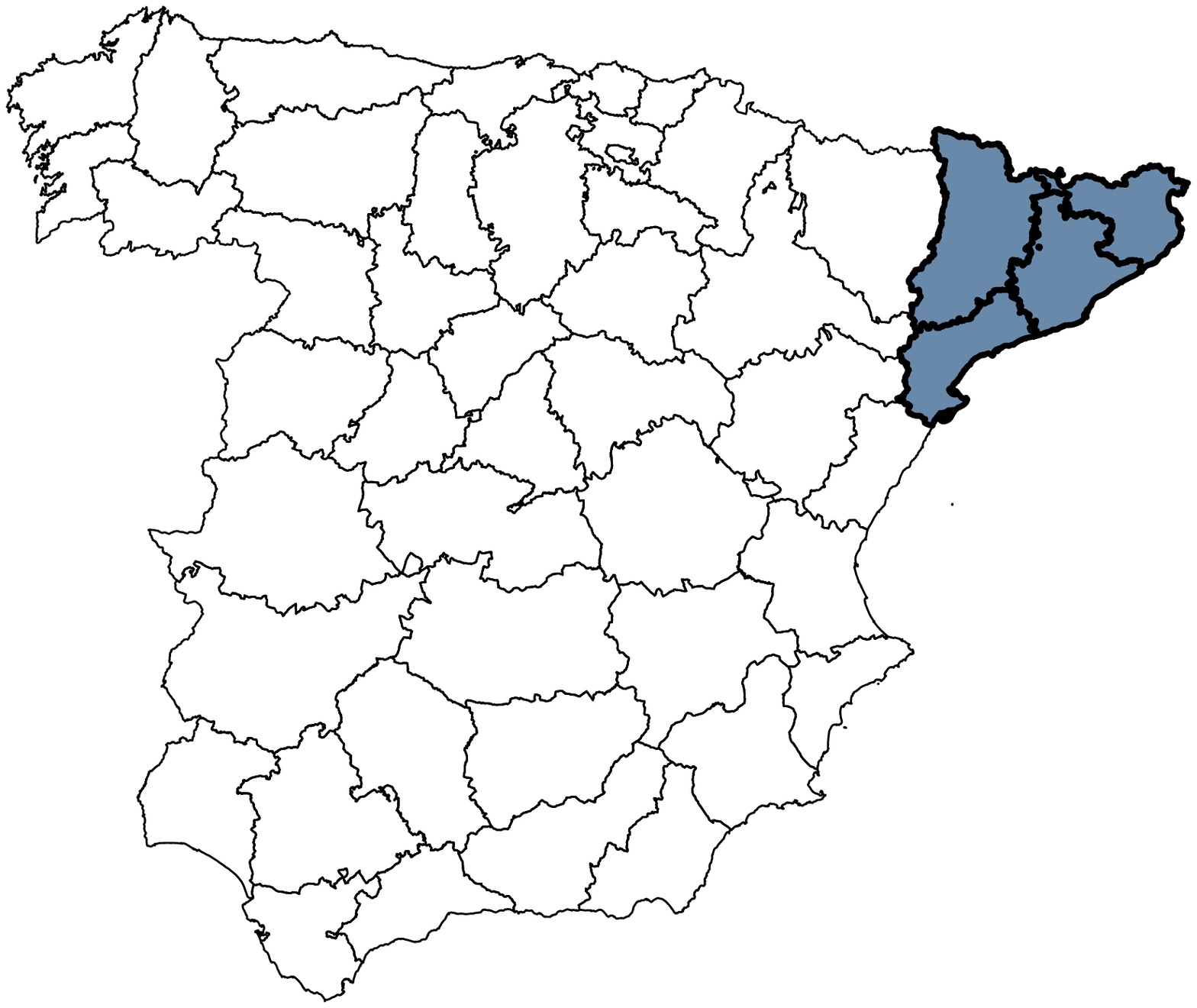}\label{fig:StudyArea_a}}
  \subfloat[]{\includegraphics[width=5cm,angle=-90]{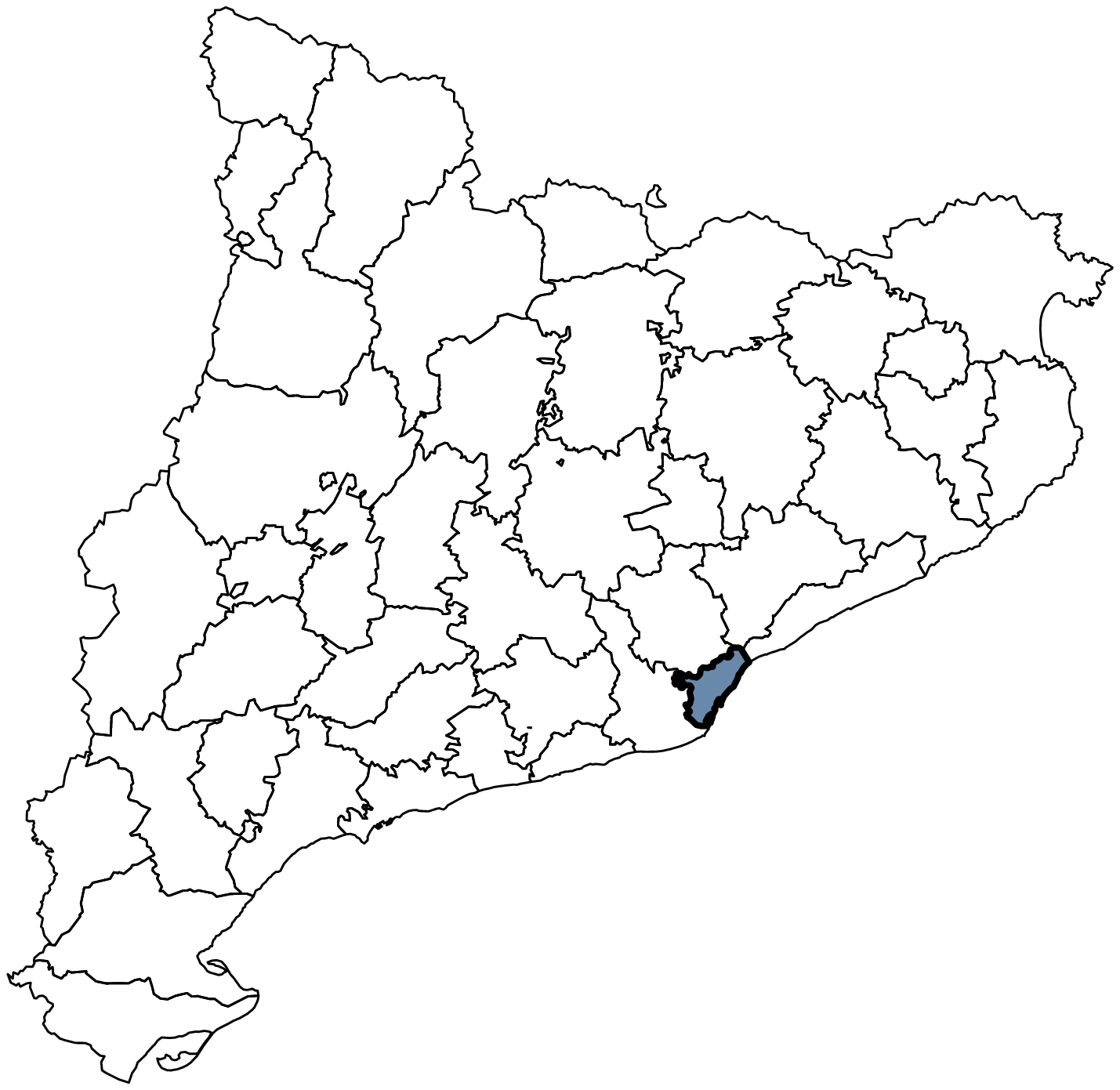}\label{fig:StudyArea_b}}\\
 \caption{Map of Spain at the province level (a) and map of Catalonia at the region level (b). In (a), the four provinces of Catalonia are highlighted. In (b), the region of Barcelonès, where the capital city of Catalonia (Barcelona) is located, is also highlighted}
 \label{fig:StudyArea} 
\end{figure}

The study has focused on Catalonia, one of the 17 Autonomous Communities of Spain. Concretely, the analysis has been carried out at the region (\textit{comarca}) level, which represents an intermediate spatial aggregation level between the province level and the city level. Thus, Catalonia is divided into 42 regions which contain about 1000 municipalities for a total of 7619494 inhabitants (as of 2019). The population sizes of these regions vary from more than 2 million people in the case of Barcelonès (which nearly represents the 30$\%$ of the population of Catalonia) to less than 4000 in the case of Alta Ribagorça. Figure \ref{fig:StudyArea} shows the location of Catalonia within Spain (Figure \ref{fig:StudyArea_a}) and a map of Catalonia at the region level (Figure \ref{fig:StudyArea_b}). 

\subsection{COVID-19 data}

A dataset containing the new daily COVID-19 cases recorded in each municipality of Catalonia, Spain, from 25 February 2020 to 24 August 2020 (covering 182 days within a total of 27 weeks) was downloaded from Catalonia's Open Data platform (\url{https://analisi.transparenciacatalunya.cat/en/}). In this dataset, cases are disaggregated according to the type of diagnostic test employed for their determination: antibody test, polymerase chain reaction (PCR) test, and serology test. Among these diagnostic methods, the PCR test has been by far the most used in Catalonia since the beginning of the pandemic. In fact, about 90\% of the cases detected up to August 24th were identified by PCR, according to the dataset downloaded. For this reason, to conduct this study, the number of daily COVID-19 cases determined by a PCR test has been considered as the response or dependent variable of the analysis.

\subsection{Environmental data}

Environmental data for the period under study has been downloaded from the OpenData platform of the State Meteorological Agency (AEMET) of Spain. Specifically, daily solar exposure (in terms of the number of hours over irradiance threshold of 120 W/m$^2$), mean temperature (in \si{\degree}C), and wind speed (in km/h) values measured from February to August 2020 by a total of 172 automatic weather stations installed all over Spain have been collected.

In order to analyse the association between the number of COVID-19 cases and the environmental conditions in the regions of Catalonia during the study period, a region-level estimation of the three environmental variables was performed for each day within the period. First, ordinary kriging \citep{cressie1988spatial} was used to estimate the daily values of the three environmental variables on a grid of points (defined at a distance of 5 km from each other) covering the whole area under study. Hence, only the stations from Catalonia and the two Autonomous Communities of Spain sharing a border with Catalonia (Aragón and the Valencian Community) may have influenced these estimates. Secondly, region-level daily estimates of the variables of interest were obtained as the average of the estimates corresponding to the points of the grid lying within the region. 

\section{Methodology}

\subsection{Software}

The R programming language \citep{teamR} has been used to carry out the present study. In particular, the R packages  \textsf{automap} \citep{automap}, \textsf{ggplot2} \citep{ggplot2}, \textsf{gstat} \citep{gstat,gstat2}, \textsf{INLA} \citep{INLA1,INLA2}, \textsf{rgdal} \citep{rgdal}, and \textsf{spdep} \citep{bivand2008applied} have been required at some points of the analysis. 

\subsection{Statistical models}

In this subsection, the different statistical models that have been considered for the analysis are described in order of complexity (from the simplest to the most complex). The precise specification of these models according to the set of terms and coefficients involved in each of them is provided in Table \ref{Models}.

\subsubsection{Basic models}

The number of new daily COVID-19 cases observed in region $i$ ($i=1,...,42$) on day $t$ ($t=1,...,182$), denoted by $O_{it}$, was assumed to follow a Poisson distribution with mean $\eta_{it}=E_{it}r_{it}$, where $E_{it}$ (offset term of the model) denotes the number of expected cases in region $i$ on day $t$, and $r_{it}$ the relative risk for region $i$ and day $t$. $E_{it}$ was calculated as a product of the total number of cases observed in Catalonia on day $t$ by the fraction of the population of Catalonia that region $i$ represents. 

The first model that was tested (Model 1) only included the fixed effect of each of the three environmental variables considered for the analysis: solar exposure ($x_{1}$), temperature ($x_2$), and wind speed ($x_{3}$). Next, a non-environmental variable such as the population density ($x_{4}$) was incorporated into the model (Model 2). For the remaining models (Models 3 to 12), a spatio-temporal approach was followed, which seems the most appropriate to account for the structure of the data under analysis. 

\subsubsection{Spatio-temporal models}

Several spatio-temporal models of increasing complexity were fitted to the data. First, a spatio-temporal model without interaction, that is, where regional and temporal effects act separately, was considered in Models 3 and 4. To model the spatial effects ($u_{i}$ and $v_{i}$), the Besag–York–Molliè (BYM) model was followed \citep{besag1991bayesian}. On the one hand, under the BYM model it is assumed that the conditional distribution of the spatially-structured effect on region $i$, $u_{i}$, is
$$u_{i}|u_{j \neq i} \sim Normal\bigg(\frac{1}{N_{i}}\sum_{j=1}^{n} w_{ij}u_{j},\frac{\sigma^2_{u}}{N_{i}}\bigg)$$
where $N_{i}$ is the number of neighbours that region $i$ has (two regions are neighbours if they are spatially contiguous), $w_{ij}$ is the element ($i$,$j$) of the row-standardised matrix of dimension $42\times42$ that represents the neighbourhood matrix for the regions ($w_{ij}=1/N_{i}$ if regions $i$ and $j$ are neighbours, otherwise $w_{ij}=0$), and $\sigma^2_{u}$ represents the variance of the spatially-structured effect. On the other hand, for the spatially-unstructured effect over the regions, denoted by $v_{i}$, an independent and identically distributed Gaussian prior is considered
$$v_{i} \sim Normal(0,\sigma^2_{v})$$
where $\sigma^2_{v}$ represents the variance of the spatially-unstructured effect of the model. 

With regard to the two temporal effects, the temporally-structured effect, $\gamma_{t}$, was modelled through a second-order random walk
$$\gamma_{t}|\gamma_{t-1},\gamma_{t-2} \sim Normal(2\gamma_{t-1}+\gamma_{t-2},\sigma^2_{\gamma})$$
where $\sigma^2_{\gamma}$ is the variance component. Finally, an independent and identically distributed Gaussian prior is chosen for $\phi_{t}$: $\phi_{t} \sim Normal(0,\sigma^2_{\phi})$.

In the case of Model 3, the random temporal effects are set on a weekly basis, whereas in Model 4 the temporal effects are set on a weekly basis. Hence, in Table \ref{Models}, the index of the temporal effects corresponding to Model 3 is actually denoted by $w(t)$ (instead of $t$), which represents the week to which the day $t$ belongs to ($w(t)=1,...,27$). The consideration of weekly effects instead of daily effects allows reducing the complexity of the model by reducing the number of parameters being involved, which reduces the chance of overfitting issues.

Then, several spatio-temporal models accounting for the presence of space-time interaction were also fitted (Models 5 to 12), among which the space-time interaction is accounted for on a weekly (Models 5 to 8) or daily basis (Models 9 to 12). In particular, the four spatio-temporal structures proposed by \cite{knorr2000bayesian} were used. Each of these structures consists in specifying the non-separable spatio-temporal term of the model according to a concrete combination of a structured/unstructured spatial effect with a structured/unstructured temporal effect. The combination of these effects is carried out through the Kronecker product of the two matrices that represent the spatial and temporal effect chosen, respectively. Table \ref{knorrheld} shows the four types of spatio-temporal interactions that can be considered following this approach.

The implementation of Models 1 to 12 was carried out through the Integrated Nested Laplace Approximation (INLA) method, which allows obtaining the posterior marginal distributions of the parameters involved in the model. Non-informative priors were chosen for the parameters corresponding to the fixed effects included in all of the models, whereas a $Gamma(1,5\cdot10^{-5})$ was used for the precision of the random effects implicated in Models 3 to 12 (these are the default priors provided by the \textsf{INLA} package). Further details on the implementation of these models in \textsf{INLA} can be found in the literature \citep{ugarte2014fitting,blangiardo2015spatial,gomez2020bayesian}. 

\subsubsection{Model variations}

Although comparing the 12 types of models introduced above constitutes the main part of the comparison, certain (minor) variations of them are also considered to extend the comparative analysis. First, the three environmental covariates were introduced into the model with a certain time lag with respect to the cases observed on the day $t$. Specifically, since COVID-19 has shown a mean incubation period of approximately 5 days, ranging from 2 to 14 days \citep{Nishiura2020,Rasmussen2020}, a lag of 0, 7, and 14 days was considered for the covariates (which implies replacing the covariate terms $x_{jit}$ present in all the expressions included in Table \ref{Models} by $x_{jit-7}$ or $x_{jit-14}$). Second, the possibility of considering the environmental covariates in their quadratic or cubic form in order to capture non-linear effects is also considered for some of the models. In these cases, the new models will be referred to only as specific modifications of the Models 1 to 12 described in Table \ref{Models}, which are those that define the fundamental modelling structures being compared.

\begin{table}[]
\centering
\begin{tabular}{c|c}
\hline
Model & $\log(r_{it})$ \\ \hline
Model 1                   &    $\mu+\log(E_{it})+\sum_{j=1}^{3}\beta_{j}x_{jit}$     \\
Model 2                   &    $\mu+\log(E_{it})+\sum_{j=1}^{4}\beta_{j}x_{jit}$     \\
Model 3                   &    $\mu+\log(E_{it})+\sum_{j=1}^{4}\beta_{j}x_{jit}+u_{i}+v_{i}+\gamma_{w(t)}+\phi_{w(t)}$    \\
Model 4                   &    $\mu+\log(E_{it})+\sum_{j=1}^{4}\beta_{j}x_{jit}+u_{i}+v_{i}+\gamma_{t}+\phi_{t}$    \\
Model 5                   &     $\mu+\log(E_{it})+\sum_{j=1}^{4}\beta_{j}x_{jit}+u_{i}+v_{i}+\gamma_{w(t)}+\phi_{w(t)}+\delta_{iw(t)}$ (I)   \\
Model 6                   &    $\mu+\log(E_{it})+\sum_{j=1}^{4}\beta_{j}x_{jit}+u_{i}+v_{i}+\gamma_{w(t)}+\phi_{w(t)}+\delta_{iw(t)}$ (II)  \\
Model 7                   &    $\mu+\log(E_{it})+\sum_{j=1}^{4}\beta_{j}x_{jit}+u_{i}+v_{i}+\gamma_{w(t)}+\phi_{w(t)}+\delta_{iw(t)}$ (III)   \\
Model 8                   &    $\mu+\log(E_{it})+\sum_{j=1}^{4}\beta_{j}x_{jit}+u_{i}+v_{i}+\gamma_{w(t)}+\phi_{w(t)}+\delta_{iw(t)}$ (IV)   \\
Model 9                   &     $\mu+\log(E_{it})+\sum_{j=1}^{4}\beta_{j}x_{jit}+u_{i}+v_{i}+\gamma_{t}+\phi_{t}+\delta_{it}$ (I)   \\
Model 10                   &    $\mu+\log(E_{it})+\sum_{j=1}^{4}\beta_{j}x_{jit}+u_{i}+v_{i}+\gamma_{t}+\phi_{t}+\delta_{it}$ (II)  \\
Model 11                   &    $\mu+\log(E_{it})+\sum_{j=1}^{4}\beta_{j}x_{jit}+u_{i}+v_{i}+\gamma_{t}+\phi_{t}+\delta_{it}$ (III)   \\
Model 12                   &    $\mu+\log(E_{it})+\sum_{j=1}^{4}\beta_{j}x_{jit}+u_{i}+v_{i}+\gamma_{t}+\phi_{t}+\delta_{it}$ (IV)   \\\hline
\end{tabular}
\caption{Description of the 12 main models that were considered for the comparison in terms of the specification of the logarithm of the relative risk, $\log(r_{it})$, corresponding to region $i$ ($i=1,...,42$) on day $t$ ($t=1,...,182$). For all the models, $\mu$ denotes the intercept of the model, $E$ the number of expected cases, and $x_{j}$ ($j=1,...,4$) the covariates. In addition, $u$ and $v$ represent the structured and unstructured random spatial effect of the model, $\gamma$ and $\phi$ the structured and unstructured random temporal effect, and $\delta$ the random spatio-temporal effect. The symbols I, II, III, IV denote the type of spatio-temporal interaction (for either $\delta_{iw(t)}$ or $\delta_{it}$) considered in Models 5 to 12, according to Table \ref{knorrheld}}
\label{Models}
\end{table}

\begin{table}[]
\centering
\begin{tabular}{c|c}
\hline
Type of spatio-temporal interaction & $R_{\delta}$ \\ \hline
 I                      &    $I_{s} \otimes I_{t}$    \\
 II                     &    $I_{s} \otimes R_{t}$    \\
 III                    &    $R_{s} \otimes I_{t}$    \\
 IV                     &    $R_{s} \otimes R_{t}$    \\ \hline
\end{tabular}
\caption{Specification of the four types of spatio-temporal interaction considered in terms of the Kronecker product of the two matrices representing the structure of the spatial and temporal effect, respectively. The matrix $I_{s}$ ($I_{t}$) represents the identity matrix, which corresponds to the unstructured spatial (temporal) effect, whereas $R_{s}$ ($R_{t}$) represents a non-identity matrix that corresponds to a specific structured spatial (temporal) effect}
\label{knorrheld}
\end{table}

\section{Results}

This section summarises the results provided by each of the statistical models fitted. First, the quality of the models is assessed. Second, the coefficients associated with the three environmental variables involved in the analysis are compared across models. Finally, the spatio-temporal effects estimated through Models 3 to 12 are described and shown graphically. 

\subsection{Model quality}

The goodness-of-fit of the models was assessed through the Deviance Information Criterion (DIC) introduced by \cite{spiegelhalter2002bayesian}. As a general rule, the model with the smallest DIC value is the best model available, as it represents the best balance between the deviance of the model and the number of parameters involved.

Thus, Model 9 including random temporal effects at the daily level and a type I spatio-temporal interaction (unstructured in space and time) showed the greatest performance in terms of the DIC (Table \ref{DICs}), while Model 11 including a type III spatio-temporal interaction (structured in space but unstructured in time) presented the second-lowest DIC. However, Model 12, which considers a type IV space-time interaction (structured in both space and time) presented an unstable behaviour and an unreliable DIC value (extremely high in comparison with the rest of the models). Among the models considering a random temporal effect weekly (Models 5 to 8), the model with the type I spatio-temporal interaction (Model 5) also presented the greatest performance according to the DIC, closely followed by Model 8 (type IV interaction).

\begin{table}[ht]
\centering
\begin{tabular}{ccccccc}
 \hline
 \multirow{3}{*}{Model} & \multicolumn{6}{c}{Lagged effect on the covariates (in days)} \\ \cline{2-7}
 & \multicolumn{2}{c}{0} & \multicolumn{2}{c}{7} & \multicolumn{2}{c}{14} \\ 
  \cline{2-7}
      & DIC & $p_{D}$ & DIC & $p_{D}$ & DIC & $p_{D}$ \\ 
  \hline
Model 1 & 72565.19 & 4.62 & 73011.11 & 4.62 & 73122.80 & 4.62 \\ 
Model 2 & 67300.15 & 5.62 & 67327.76 & 5.62 & 67317.83 & 5.62 \\ 
Model 3 & 52467.55 & 69.88 & 52445.03 & 69.91 & 52471.95 & 69.86 \\ 
Model 4 & 52440.54 & 88.46 & 52421.46 & 90.89 & 52459.30 & 87.80 \\ 
Model 5 & 31851.36 & 805.13 & 31858.83 & 804.78 & 31855.27 & 804.92 \\ 
Model 6 & 32060.54 & 584.64 & 32073.40 & 584.25 & 32067.71 & 584.21 \\ 
Model 7 & 31991.11 & 763.45 & 32001.95 & 762.70 & 31998.69 & 763.53 \\ 
Model 8 & 31860.42 & 573.30 & 31868.50 & 573.71 & 31864.40 & 572.79 \\ 
Model 9 & 26162.89 & 3510.99 & 26163.56 & 3511.21 & 26161.46 & 3513.11 \\ 
Model 10 & 29175.52 & 1239.85 & 29197.48 & 1237.62 & 29192.74 & 1228.14 \\ 
Model 11 & 26240.72 & 3286.78 & 26245.35 & 3286.84 & 26246.03 & 3285.53 \\
Model 12 & - & - & - & - & - & - \\ 
   \hline
\end{tabular}
\caption{DIC values and effective number of parameters ($p_{D}$) corresponding to Models 1 to 12, considering a lagged effect on the covariates of 0, 7, or 14 days. In the case of Model 12, the values obtained for the DIC and the $p_{D}$ were not comparable to those of the rest of models (extremely high), so they are omitted (-)}
\label{DICs}
\end{table}

\subsection{Environmental effects}

\begin{figure}
    \centering
    \includegraphics[width=0.6\textwidth,angle=-90]{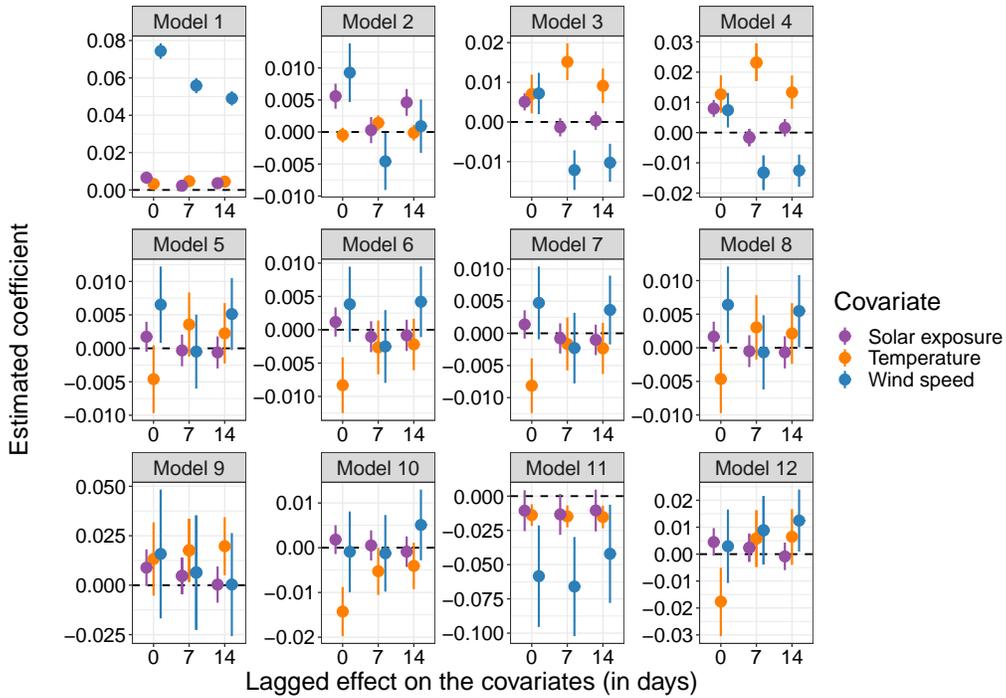}
    \caption{Summary of the estimates obtained for the coefficients associated with environmental covariates for each of the 12 models fitted, considering a lagged effect on the covariates of 0, 7, or 14 days}
    \label{fig:CovariateEffects}
\end{figure}

Regarding the effect of each environmental covariate on the spread of COVID-19, the main conclusion would be that the choice of the model has a strong impact on the results, as shown in Figure \ref{fig:CovariateEffects}. First, Model 1 suggests that solar exposure, wind speed, and temperature have a positive association with COVID-19 spread. Besides, these associations are consistent across the different time lags (0, 7, and 14 days) explored for the covariates. A visible characteristic of Model 1 is the narrowness of the confidence intervals associated with the mean estimates of the effects, which causes that all the associations are statistically significant with 95\% confidence. However, the results provided by Model 2 (where population density is incorporated into the model) increase the uncertainty about the possible association between COVID-19 spread and the environment. Now, there is great inconsistency across lags for the three environmental covariates, which makes it difficult to achieve solid conclusions on their effects. Indeed, the difficulty in establishing an association between environmental covariates and the number of daily COVID-19 cases is maintained if the results provided by Models 3 to 12 are analysed. Some models suggest that there is a positive association between COVID-19 and temperature (Models 3 and 4), or a negative association between COVID-19 daily cases and wind speed (Model 11). Considering the best model in terms of the DIC, Model 9, it could be concluded that there is a statistically significant positive association between mean temperature and daily new cases, and a non-significant association with solar exposure and wind speed. However, the modification of Model 9 through the addition of the three environmental covariates in its quadratic and cubic form ($x_{i}^{2}$ and $x_{i}^{3}$, $i=1,...,3$), even though it reduces the DIC of Model 9 to 26106.22, does not reveal any significant association between these covariates and the daily number of COVID-19 cases. Therefore, there seems to be too much uncertainty to establish an association between these environmental conditions and the development of the pandemic, on the basis of the data examined in this study.

%\begin{figure}[ht] 
%  \centering
%  \subfloat[]{\includegraphics[width=3.3cm,angle=-90]{NonLinearEffects_Model9B_Sun.eps}\label{fig:NonLinearEffects_a}}
%  \subfloat[]{\includegraphics[width=3.3cm,angle=-90]{NonLinearEffects_Model9B_Temp.eps}\label{fig:NonLinearEffects_b}}
%  \subfloat[]{\includegraphics[width=3.3cm,angle=-90]{NonLinearEffects_Model9B_Wind.eps}\label{fig:NonLinearEffects_c}}\\
% \caption{}
% \label{fig:NonLinearEffects} 
%\end{figure}

\subsection{Spatio-temporal effects}

The inclusion of spatio-temporal effects helps to understand how the disease has spread throughout the territory under study. Specifically, the estimates of random spatial and temporal effects and their interaction allow assigning a relative risk to each spatial, temporal, or spatio-temporal unit under analysis. These relative risks are obtained by exponentiating the space-time parameters that describe the log($r_{it}$) expression in each of the models. In the remainder of the section, the estimates of the random spatial, temporal and spatio-temporal effects that are displayed are those that correspond to the models that consider a 7-day lagged effect in the environmental covariates (because this seems the most reasonable choice considering the already known incubation time). 

\begin{figure}[ht] 
  \centering
  \subfloat[]{\includegraphics[width=5cm,angle=-90]{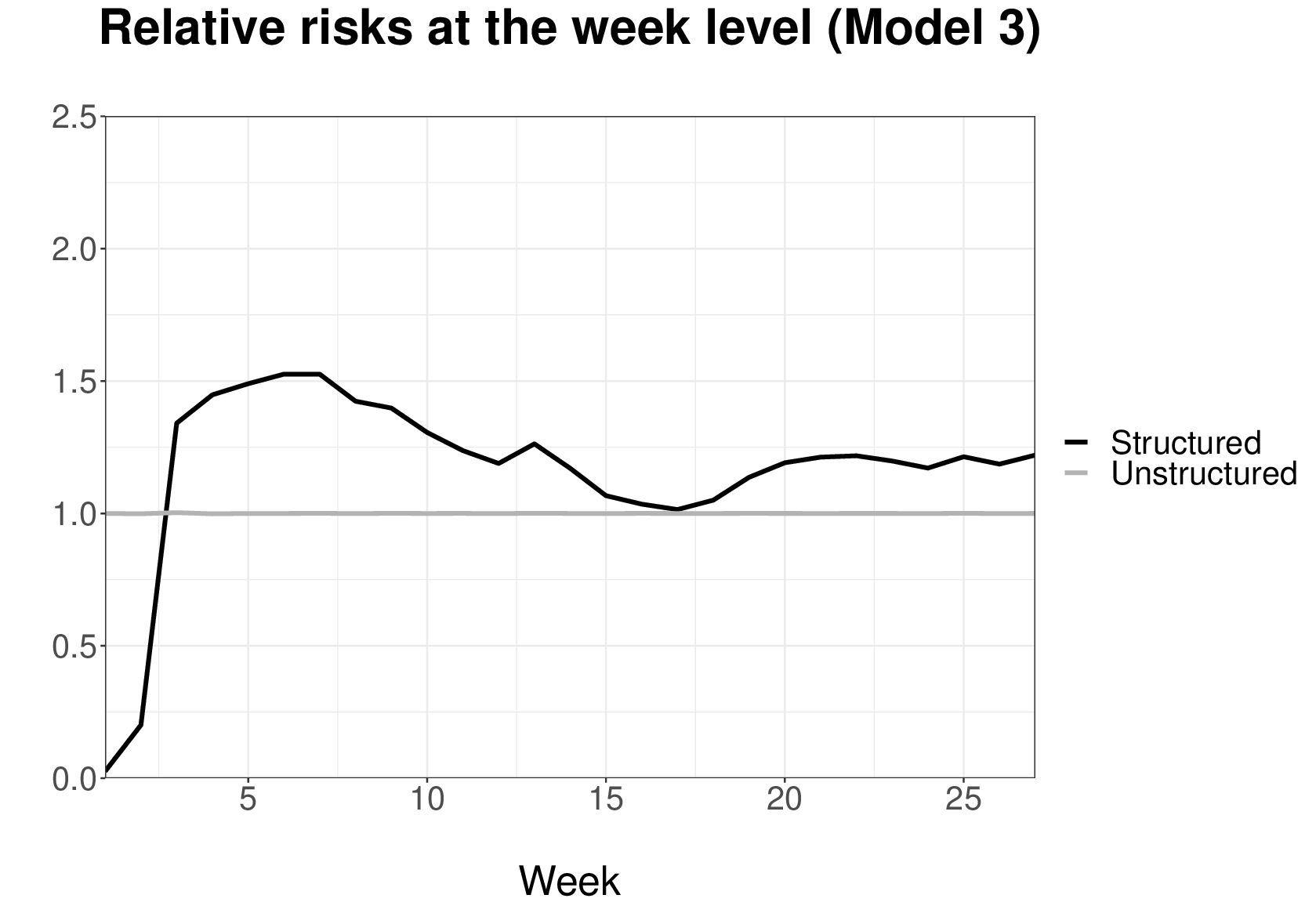}\label{fig:TemporalEffects_a}}
  \subfloat[]{\includegraphics[width=5cm,angle=-90]{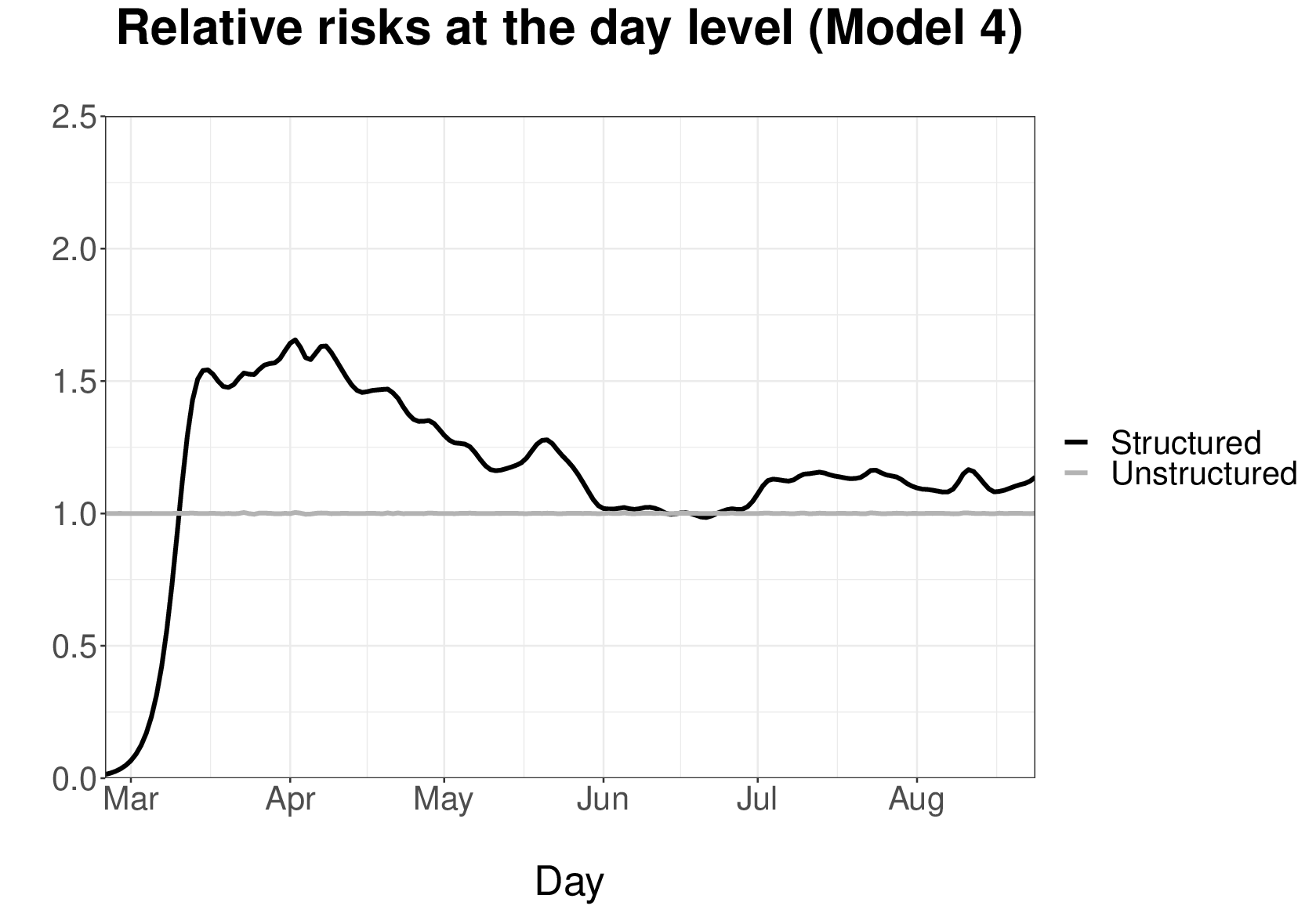}\label{fig:TemporalEffects_b}}\\
 \caption{Relative risks on a weekly and a daily basis according to the structured and unstructured temporal random effects estimated through Models 3 (a) and 4 (b). The relative risk corresponding to the structured component is computed as either $\exp(\gamma_{w(t)})$ or $\exp(\gamma_{t})$, whereas the one corresponding to the unstructured component is computed as either $\exp(\phi_{w(t)})$ or $\exp(\phi_{t})$}
 \label{fig:TemporalEffects} 
\end{figure}

Hence, Figure \ref{fig:TemporalEffects} shows relative risks over time
in terms of the random temporal effects estimated through Models 3 (including weekly effects) and 4 (including daily effects). The relative risk represented by the structured component of the random temporal effect (either $\exp(\gamma_{w(t)})$ in Model 3 or $\exp(\gamma_{t})$ in Model 4, for $t=1,...,182$) captures the evolution of the pandemic in Catalonia: the relative risk was 0 at the beginning of March 2020, reached a peak in April, and then decreased for the following months until July, when it started to increase again. Oppositely, the relative risk associated with the unstructured component (either $\exp(\phi_{w(t)})$ in Model 3 or $\exp(\phi_{t})$ in Model 4) barely fluctuates around 1, which suggests that there were not notable overall changes in the relative risk during the period of study that were solely attributable to single days within the period.

\begin{figure}[ht] 
  \centering
  \subfloat[]{\includegraphics[width=5cm,angle=-90]{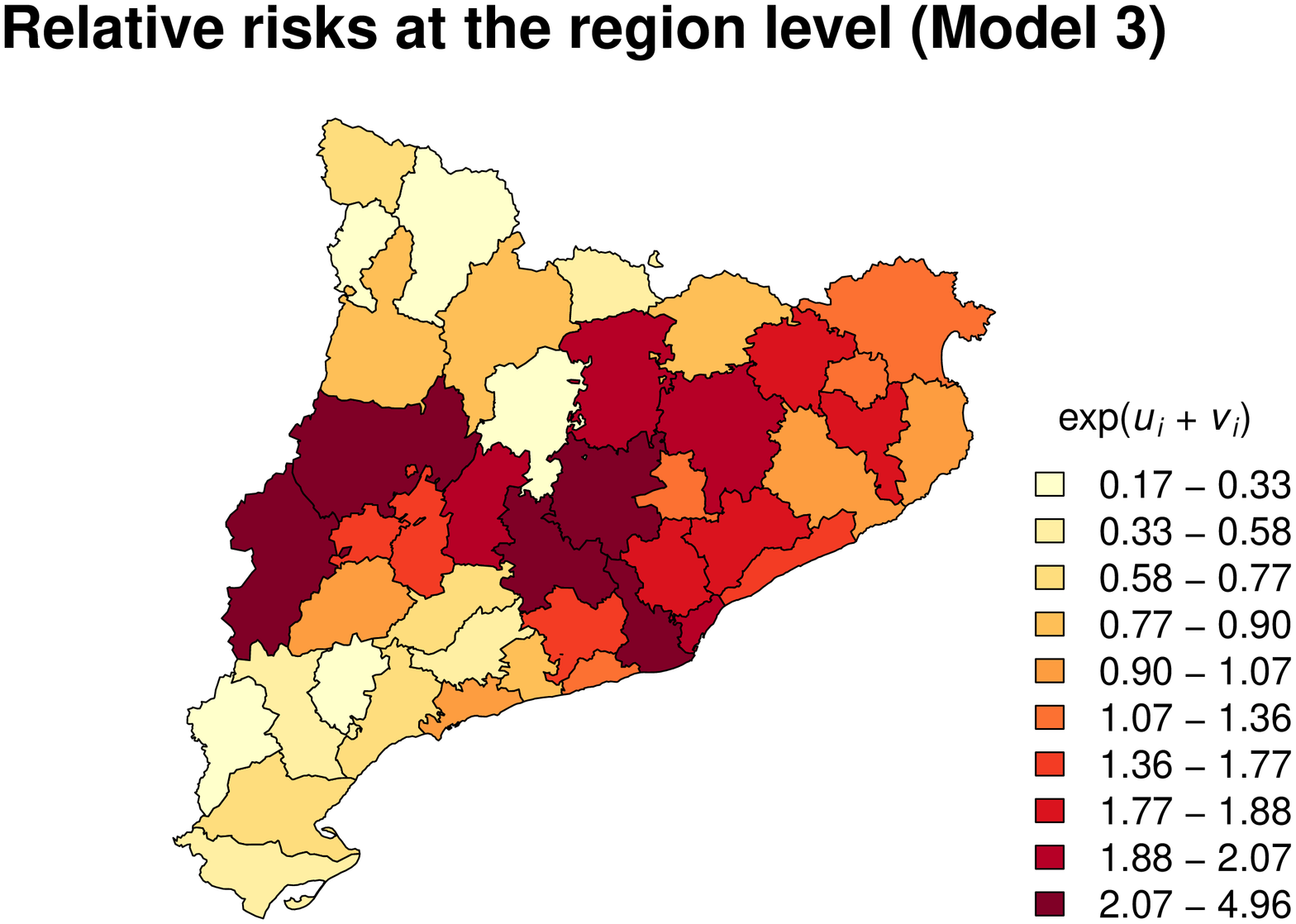}\label{fig:SpatialEffects_a}}
  \subfloat[]{\includegraphics[width=5cm,angle=-90]{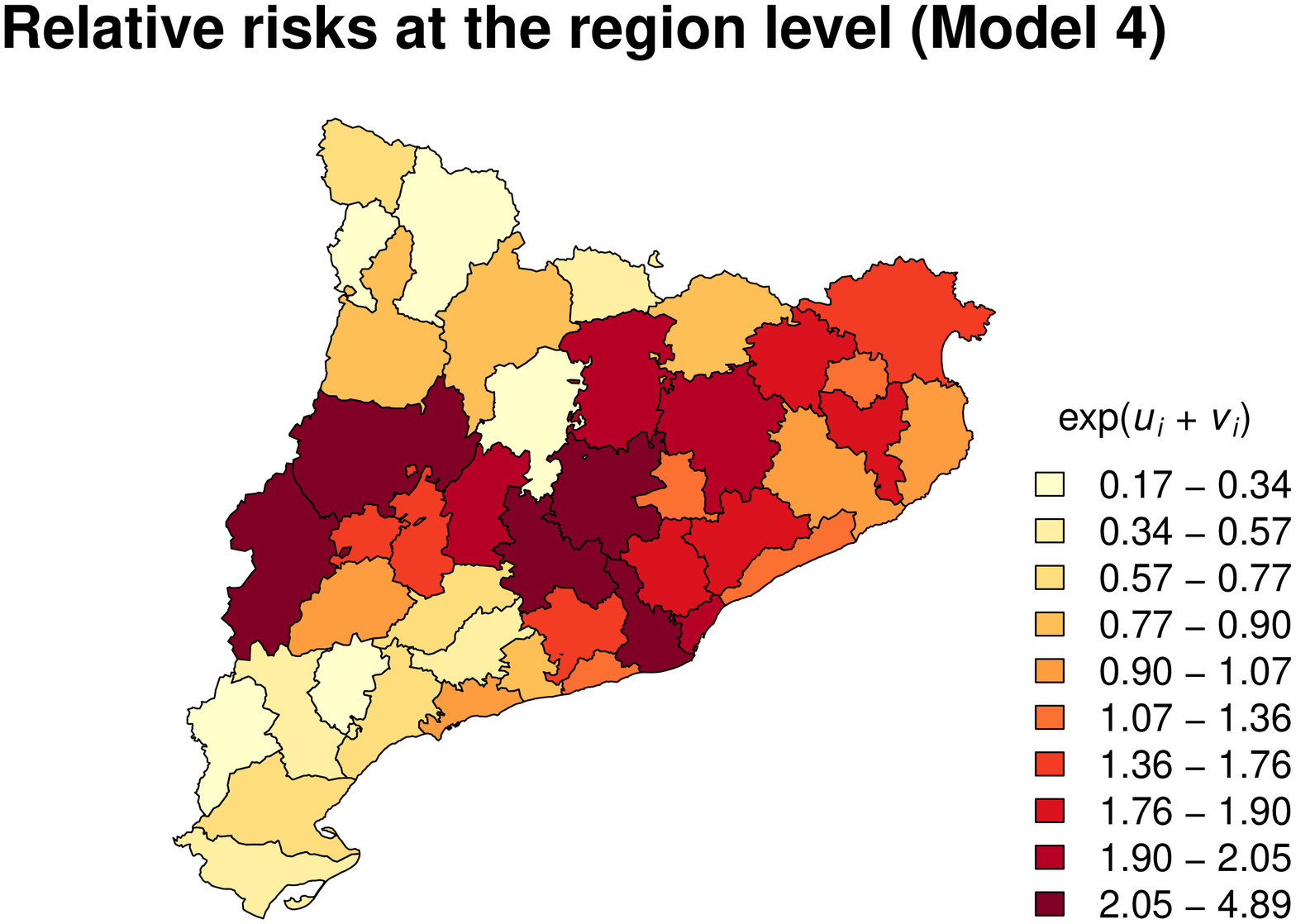}\label{fig:SpatialEffects_b}}\\
 \caption{Global relative risks at the region level estimated for the period under study (computed as $\exp(u_{i}+v_{i})$) considering Model 3 (a) and Model 4 (b)}
 \label{fig:SpatialEffects} 
\end{figure}

With regard to the random spatial effects, Figure \ref{fig:SpatialEffects} displays the values of $\exp(u_{i}+v_{i})$ corresponding to Model 3 (although the differences between the two models are almost negligible). It can be observed that the regions in the central zone of Catalonia, which covers from the surroundings of the Barcelonès region to some of the regions in the west of Catalonia that are bordering Aragón, experienced the highest relative risks during the period under research. One of these regions located in western Catalonia, called Segrià, reached the highest relative risk for the period under consideration, presenting a relative risk very close to 5 (for both Models 3 and 4). The regions of Anoia, Bages, Baix Llobregat, Barcelonès, and Noguera were the ones that presented the closest relative risks to Segrià, although they were considerably smaller, only ranging from 2 to 3.

\begin{figure}
    \centering
    \includegraphics[width=0.65\textwidth,angle=-90]{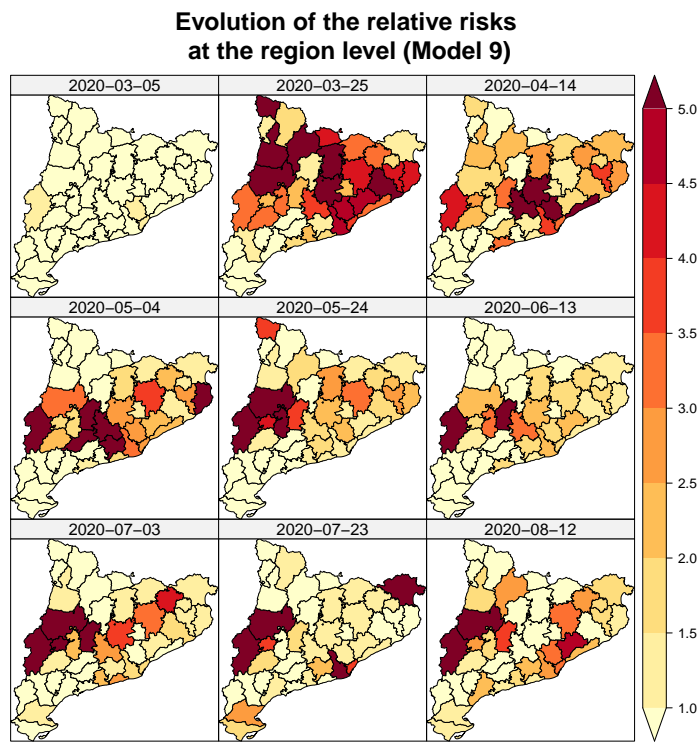}
    \caption{Relative risks at the region level estimated for a selection of days within the period under study (computed as $\exp(u_{i}+v_{i}+\gamma_{t}+\phi_{t}+\delta_{it})$) with Model 9}
    \label{fig:SpatioTemporalEffects}
\end{figure}

Finally, the spatio-temporal relative risks provided by Model 9 for a selection of days within the period of study are shown in Figure \ref{fig:SpatioTemporalEffects} (these are computed as $\exp(u_{i}+v_{i}+\gamma_{t}+\phi_{t}+\delta_{it})$). The inclusion of space-time interaction terms is essential to allow the model to capture certain variations in relative risks across both regions and subperiods. Thus, by observing the evolution of the relative risks across regions and days in Figure \ref{fig:SpatioTemporalEffects}, it can be appreciated how certain regions of the central zone of Catalonia presented quite different relative risks along time. These variations in the relative risk are overlooked if one only considers global spatial effects (Figure \ref{fig:SpatialEffects}). Concretely, the highest relative risks for most of the regions were achieved between the end of March 2020 and the beginning of April 2020. Then, for the following months, the relative risks were generally lower across entire Catalonia, except for some regions in the west of Catalonia such as Segrià, which has been presenting higher relative risks since the month of May. 

\begin{figure}
    \centering
    \includegraphics[width=0.4\textwidth,angle=-90]{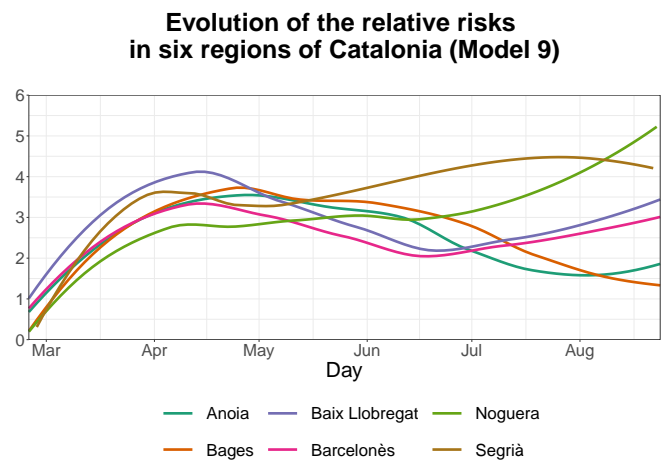}
    \caption{Evolution of the relative risks (computed as $\exp(u_{i}+v_{i}+\gamma_{t}+\phi_{t}+\delta_{it})$, according to the estimates provided by Model 9) in the six regions of Catalonia with the highest global relative risks (according to the estimates provided by Models 3 and 4). To make this plot, the relative risks provided by Model 9 have been smoothed through a locally estimated scatterplot smoothing (LOESS) regression \citep{fox2018r} for ease of visualisation and interpretation}
    \label{fig:SpatioTemporalEffectsRegions}
\end{figure}

To better appreciate the evolution of the relative risk in some highly affected regions of Catalonia, Figure \ref{fig:SpatioTemporalEffectsRegions} shows the evolution of the relative risk (according to Model 9) that correspond to a selection of regions of Catalonia (the six regions mentioned above, which presented the highest relative risk according to Models 3 and 4). It is important to note that the relative risks provided by Model 9 at the day level for each of the regions are quite erratic so that the values were previously smoothed through a locally estimated scatterplot smoothing (LOESS) regression \citep{fox2018r} to create Figure \ref{fig:SpatioTemporalEffectsRegions}. Hence, Figure \ref{fig:SpatioTemporalEffectsRegions} indicates that, except for Segrià, all these regions reached a peak in the relative risk at the beginning of April 2020, and then decreased until July, when it started growing again. This temporal pattern corresponds to the overall relative risk over time shown in Figure \ref{fig:TemporalEffects}. In the case of Segrià, however, the relative risk kept growing until August, when it started to show a slight decrease.

\section{Discussion and conclusions}

This paper has shown how the choice of a certain type of statistical model to evaluate the association between a set of covariates and a response variable can seriously influence the results. In the context of the study of the association between the evolution of the COVID-19 pandemic and environmental conditions, this fact seems to be occurring remarkably. In this regard, the lack of consideration of certain non-environmental variables, and overlooking spatio-temporal effects appear inadequate.

Furthermore, many other methodological choices could have some influence on the association between the environment and the COVID-19 propagation that have not been considered. For instance, the definition of different neighbourhood relationships between the regions, the consideration of more non-environmental covariates such as the inter-region mobility or the age structure of the population, and the selection of the most suitable spatio-temporal unit for the analysis, which implies dealing with the modifiable areal unit problem \citep[MAUP;][]{openshaw1981quantitative} and the modifiable temporal unit problem \citep[MTUP;][]{cheng2014modifiable}, are other issues that deserve attention. In particular, concerning the MAUP, it is important to note that some geographical units at the subregional level (such as cities or even city districts) may present certain unique characteristics that require consideration for performing an accurate analysis of the evolution of the pandemic. For instance, \cite{wang2020modifiable} recently found that the association between COVID-19 mortality and NO$_{2}$ levels depends on the aggregation level (considering four different spatial aggregations, including cities and provinces), which indicates the presence of the MAUP. 

Another important aspect to consider is the fact that the detection rate of cases remains far from 100\% since the beginning of the COVID-19 pandemic. In particular, if detection rates vary spatially, this could have a notable impact on the results. For instance, in the case of Spain, differences in detection rates between geographical units belonging to different Autonomous Communities are likely to arise because the competences in health policy and organisation are established at this territorial level. To mitigate this problem, the existence of seroepidemiological studies that provide estimates of the prevalence of COVID-19 at the province level \citep{pollan2020prevalence}, or the availability of reliable COVID-19 mortality data \citep{langousis2020undersampling} could be helpful. 

In conclusion, it seems clear that the data modelling approach can have a strong impact on the conclusions that can be drawn from it. For this reason, in the specific case of the ongoing line of research that focuses on unveiling the effects of the environment on the spread of COVID-19, the employment of models that properly take into account the structure of the data, the consideration of non-environmental variables, or the performance of sensitivity analyses on the results, seem highly-advisable strategies to avoid the persistence of highly contradictory results which could make decision-making against the COVID-19 pandemic even more difficult.

\normalsize
\clearpage
\bibliography{bibliography}

\end{document}